\documentclass[journal]{IEEEtran}
\usepackage{cite}     
\usepackage{graphicx}
\usepackage{color}

\usepackage{amsmath}

\begin{document}
\title{Energy Reconstruction of Hadron Showers in the CALICE Calorimeters}

\author{{Frank Simon \\ {\small{Max-Planck-Institut f\"ur Physik, Munich, Germany and Excellence Cluster Universe, Technical University Munich, Germany }}\\ \vspace{3mm}{\it{on behalf of the CALICE Collaboration}}}
}

\maketitle

\begin{abstract}
The CALICE collaboration has constructed highly granular electromagnetic and hadronic calorimeter prototypes to evaluate technologies for the use in detector systems at the future International Linear Collider. These calorimeters have been tested extensively in particle beams at CERN and at Fermilab. We present analysis results for hadronic events recorded at CERN with a SiW ECAL, a scintillator tile HCAL and a scintillator strip tail catcher, the latter two with SiPM readout, focusing both on the HCAL alone and on the complete calorimeter setup.  Particular emphasis is placed on the study of the linearity of the detector response and on the single particle energy resolution. The high granularity of the detectors was used to perform first studies of software compensation based on the local shower energy density, yielding significant improvements in the energy resolution. The required calibration precision to achieve this resolution, and the effect of calibration uncertainties, for the CALICE HCAL as well as for a complete hadron calorimeter at ILC, has been studied in detail. The prospects of using minimum-ionizing track segments within hadronic showers for calibration are also discussed.
\end{abstract}


\maketitle
\thispagestyle{empty}

\section{The CALICE Detectors}

The goal of the CALICE experimental program is to establish novel technologies for calorimetry in detectors at a future linear $e^+e^-$ collider, and to record electromagnetic and hadronic shower data with unprecedented three dimensional spatial resolution for the
validation of simulation codes and for the test and development of reconstruction algorithms. Such highly granular calorimeters are necessary to achieve an unprecedented jet energy resolution at the International Linear Collider \cite{:2007sg} using particle flow algorithms \cite{Thomson:2009rp}.

\begin{figure}
\centering
\includegraphics[width=0.48\textwidth]{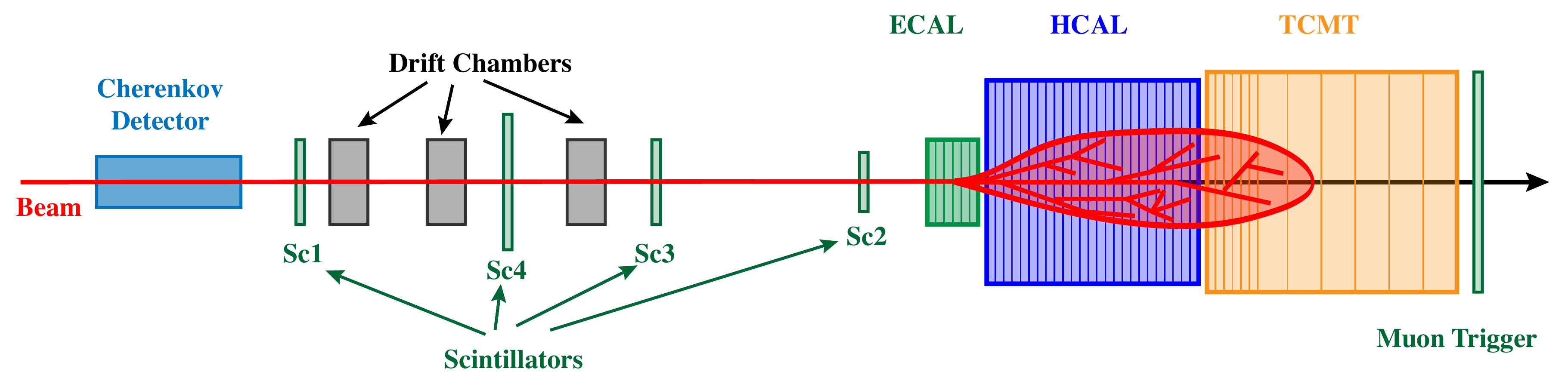}
\caption{Schematic of the CALICE experimental setup at
CERN, with electromagnetic and hadronic calorimetry as
well as a tail catcher and muon tracker downstream of the
calorimeters.}
\label{fig:CALICESetup}
\end{figure}

The CALICE test beam setup \cite{Eigen:2006eq} consists of a silicon-tungsten electromagnetic calorimeter (ECAL), an analog scintillator-steel hadron calorimeter (AHCAL) and tail catcher/muon tracker (TCMT), the latter two both with individual cell readout by silicon photomultipliers (SiPMs) \cite{Bondarenko:2000in}. This setup has been tested extensively in electron, muon and hadron beams at CERN and at the Meson Test Beam Facility at Fermilab. Figure \ref{fig:CALICESetup}  shows the schematic setup of the CALICE detectors in the CERN H6 test beam area, where data was taken in 2006 and 2007. The currently ongoing and the future program include the study of alternative technologies, such as an ECAL using scintillator strips with SiPM readout, digital hadron calorimeters with active layers based on gas detectors, and the investigation of alternative absorber materials for the hadron calorimeter for multi-TeV colliders.

The performance of the current setup for hadronic showers is driven by the analog HCAL, which is  a 38 layer sampling calorimeter with 5 mm thick scintillator layers sandwiched by 2 cm of steel. The lateral dimensions are roughly 1$\times$1 m$^2$, the total thickness amounts to 4.5 nuclear interaction lengths. The active layers are built out of scintillator tiles with sizes ranging from 30$\times$30 mm$^2$ in the core of the detector to 120$\times$120 mm$^2$. Each tile is read out by a built-in SiPM, produced by the MEPhI/PULSAR group \cite{Bondarenko:2000in}. In total, the calorimeter has 7608 channels.

In the present paper, data taken in at CERN in 2007 with the complete installation of the silicon tungsten ECAL, the analog HCAL and the TCMT, are being discussed. The total thickness of the setup amounts to more than 10 nuclear interaction lengths, guaranteeing good longitudinal containment of hadronic showers.

\section{Calibration Strategy and Requirements}

Calibrating a hadronic calorimeter with approximately 8 million channels at a future ILC detector is a significant challenge. The single photon resolution of the SiPMs, together with an LED light system, can be used to perform a gain calibration of the photon sensor, and to monitor changes due to short term variations of environmental parameters \cite{Simon:2008qj}.  

\begin{figure}[hbt]
\begin{center}
\includegraphics[width=0.40\textwidth]{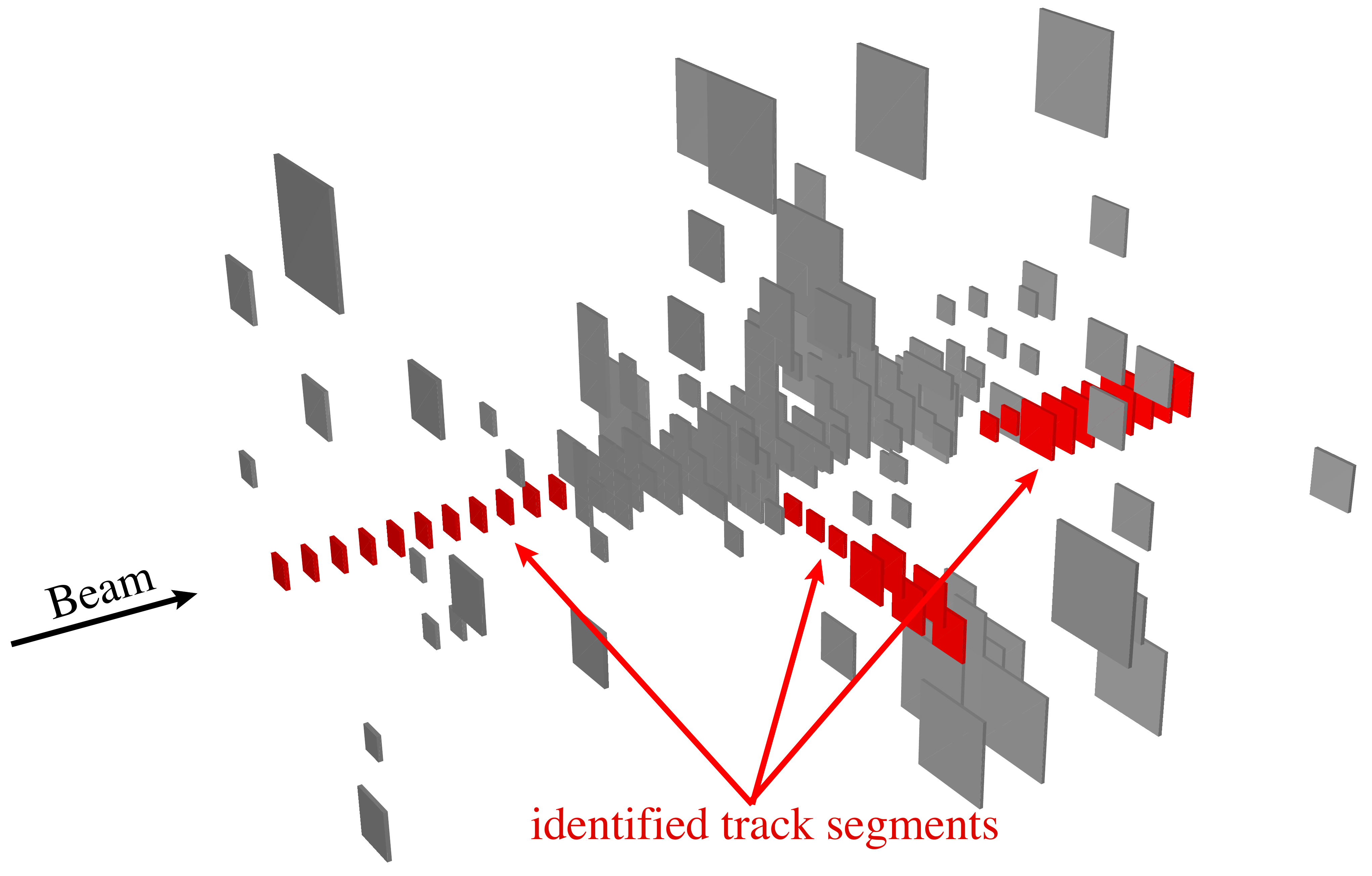}
\end{center}
\vspace{-4mm}
\caption{\label{fig:EventDisplay}Identification of minimum-ionizing track segments within hadronic showers.}
\end{figure}

For a channel-by-channel intercalibration of the complete cell response, particles are necessary. Before installation, such a calibration can be performed in a beam, analogous to the muon calibration currently used for the CALICE detector system \cite{Simon:2008qj}. 
Monitoring of long term variations, such as changes in the light yield of the scintillator, and the intercalibration of individual detector sub-modules has then to be performed in situ. For substantial parts of the calorimeter system,in particular for the non-horizontal modules of the calorimeter barrel, this cannot be achieved with cosmic or beam halo muons, in particular in underground locations. The detection rate for cosmic muons is further reduced by the power pulsing of the front-end electronics, which will be powered on for less than 1\% of the time.

An alternative calibration strategy is the use of track segments identified within hadronic showers. Since hadronic showers are typically relatively sparse, the high granularity of the hadron calorimeter allows the reconstruction of secondary hadrons that travel an appreciable distance before interacting again in the detector material. Figure \ref{fig:EventDisplay} shows the incoming primary track and two secondary tracks reconstructed in a 25 GeV pion shower in the CALICE analog HCAL. The cells along these tracks exhibit the typical energy spectrum expected for minimum-ionizing particles. The identified tracks can thus be used as a calibration tool. In the CALICE detector prototype, this technique was used to study the temperature dependence of the detector response \cite{Simon:2009mw}. 

Detailed simulations within the ILD detector concept \cite{ILD}, both at the $Z^0$ resonance and at a center of mass energy of 500 GeV, have been performed to study the applicability of this method to a full collider detector. In addition, the impact of the calibration quality in the HCAL on the overall physics performance of the experiment was studied in detail. These studies showed that sufficient statistics for an intercalibration of individual electronics subunits in the first 20 layers, and for a layer-wise calibration within calorimeter modules for layers beyond 20 is achievable with modest running times, both at the $Z^0$ and at full energy \cite{Lu:2009ze}. The running times for an in-situ cell-by-cell calibration however are prohibitive. The studies also showed that spreads of up to 10\% in the cell-to-cell and layer-to-layer calibration can be tolerated without significant impact on the physics performance \cite{Lu:2009ze}.  

The scheme of intercalibrating the cells of individual modules before installation, and then performing a module-to-module calibration in situ using track segments identified in hadronic showers has been successfully demonstrated using test beam data taken at CERN, with a calibration performed at Fermilab after a complete disassembly and reassembly of the calorimeter. The calibration constants for individual channels were transported to the conditions of the data runs taken at CERN using the known temperature and voltage dependence of the response of each channel. The overall scale of the calibration was then corrected by a layerwise correction. After this procedure, a consistent energy response and resolution was obtained for the original calibration performed in-situ at CERN and the calibration transferred from Fermilab, demonstrating the validity of the proposed calibration scheme \cite{Lu:2009ze}.

\section{Energy Reconstruction and Performance for Hadrons}

In addition to the importance of high segmentation for particle flow algorithms, the granularity of the calorimeters can also be used for software compensation procedures. Within the CALICE calorimeters, such methods can be applied to improve the single particle energy resolution. For intrinsically non-compensating calorimeters, the detector response is typically larger for electromagnetic than for hadronic showers. Since hadronic showers contain an electromagnetic component from the production of neutral pions in the cascade which fluctuates from event to event, this reduces the energy resolution and leads to non-linearities in the response.

\begin{figure}[hbt]
\begin{center}
\includegraphics[width=0.48\textwidth]{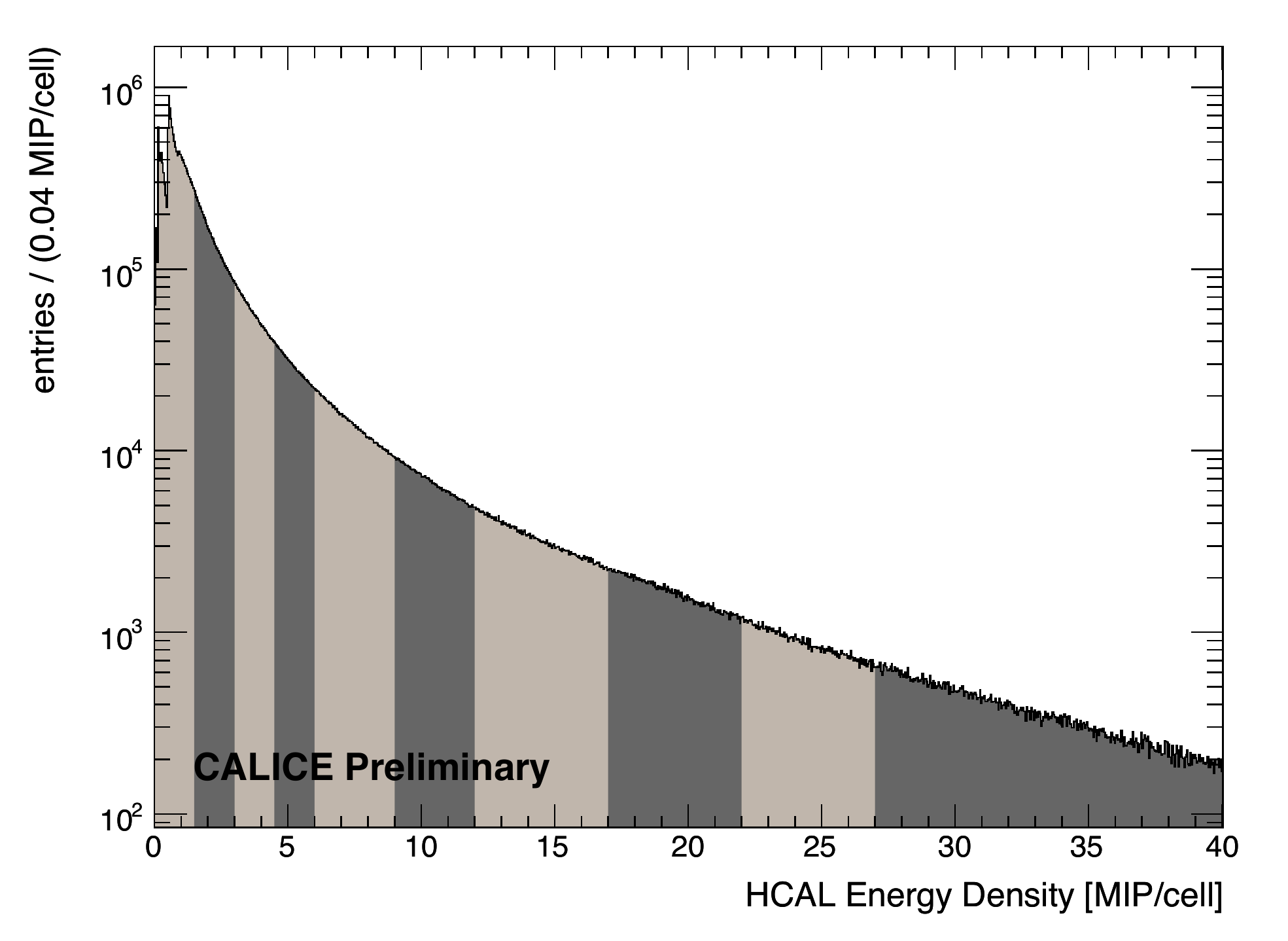}
\end{center}
\caption{\label{fig:Binning} Energy content per cell in the HCAL for 18 GeV pions, in units of minimum ionizing particle equivalents. The binning used for the weighting procedure is indicated by the shaded bands.}
\end{figure}

By identifying electromagnetic and hadronic subshowers and assigning different weights to their energy in the total reconstructed energy, resolution and linearity can be significantly improved. Since electromagnetic showers tend to be denser than purely hadronic showers, the local shower density can be used as a discriminating variable to improve the hadronic energy resolution \cite{Abramowicz:1980iv}. Depending on this energy density, the conversion factor from the raw channel signal to an energy in units of GeV is adjusted, with lower factors used in the case of higher energy densities. For an initial study, each cell in a given shower was weighted with a factor chosen according to its energy content. 

\begin{figure}[hbt]
\begin{center}
\includegraphics[width=0.44\textwidth]{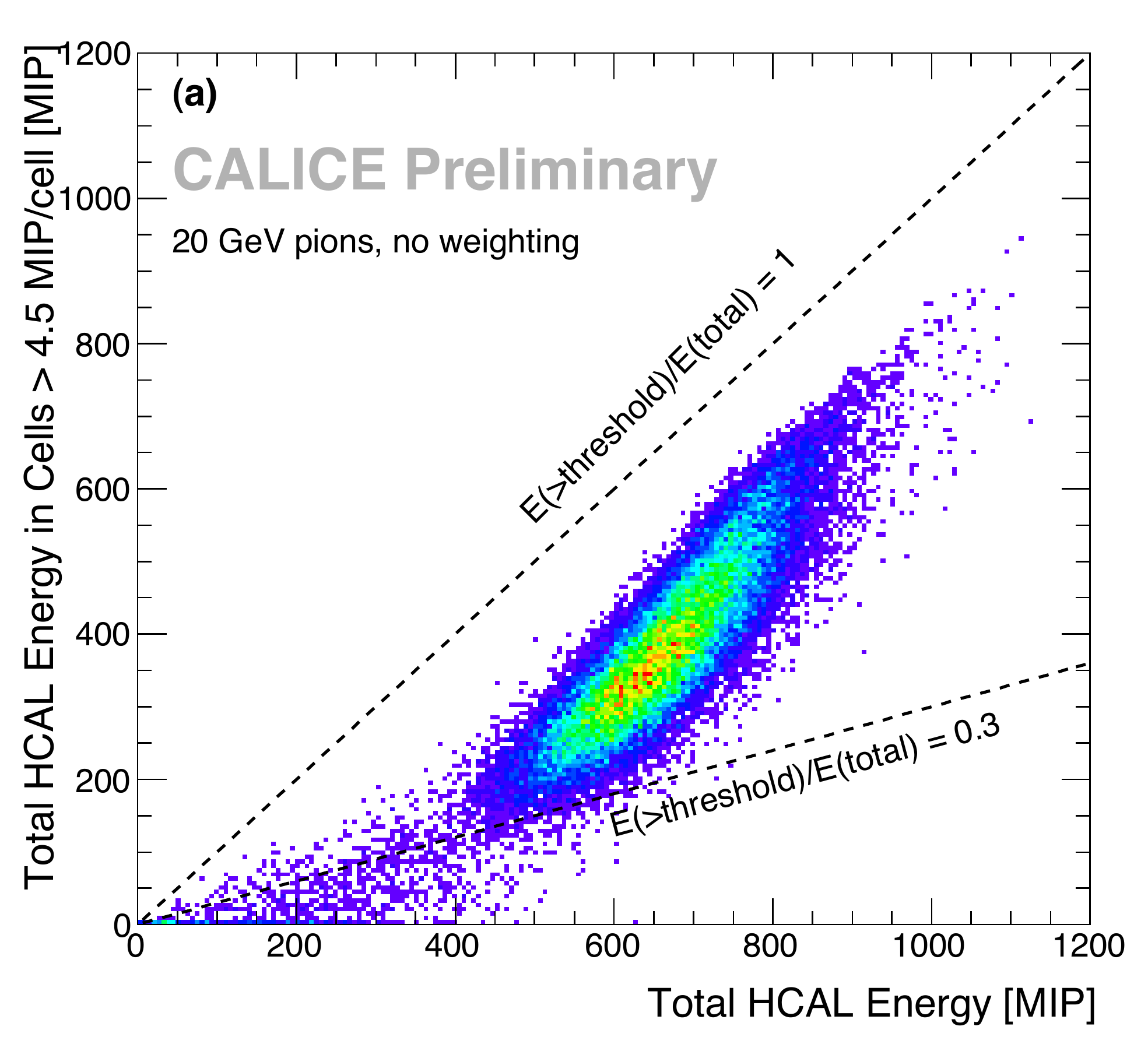}\\
\includegraphics[width=0.44\textwidth]{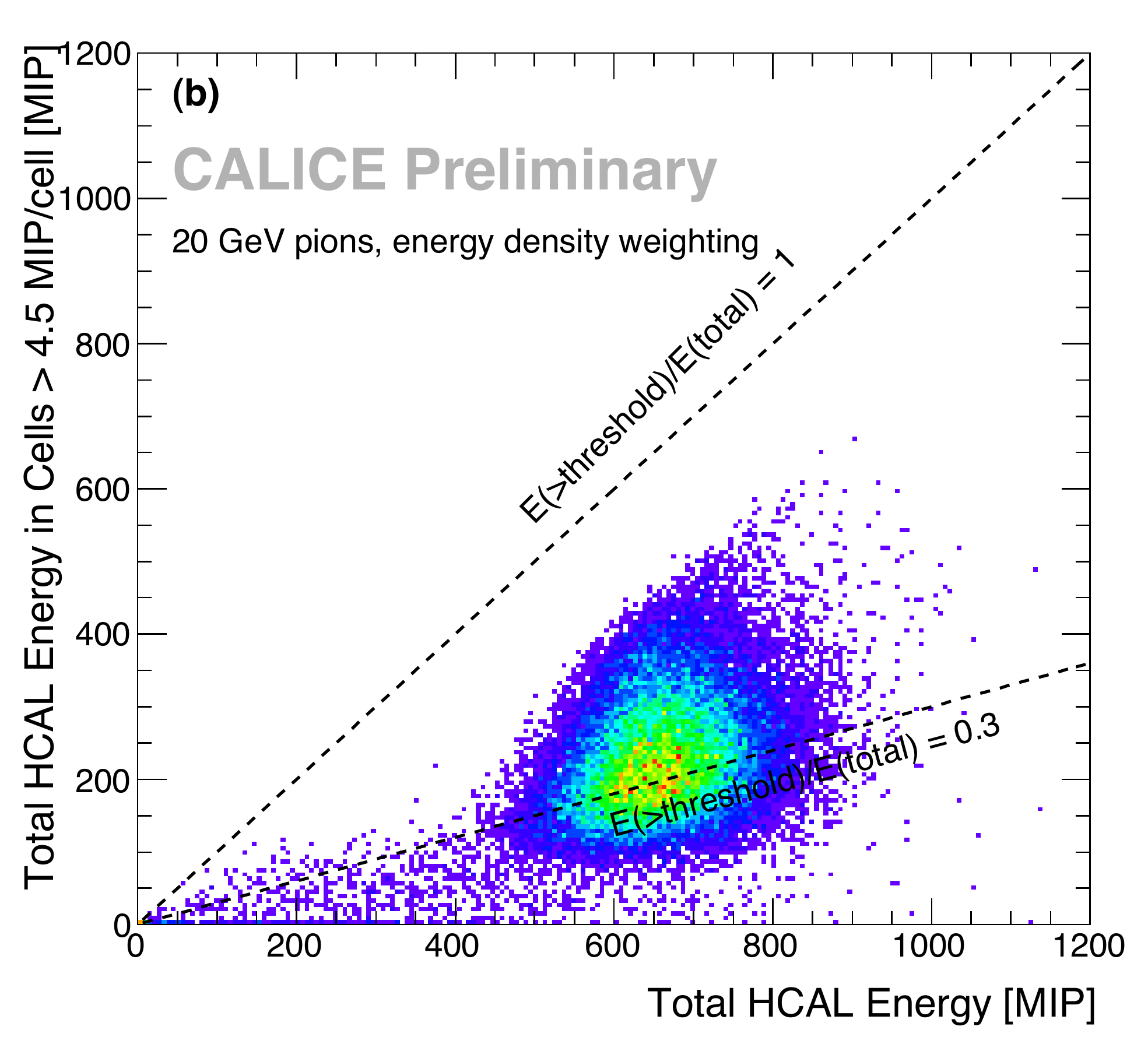}\\
\end{center}
\caption{\label{fig:Correlation} Correlation of the total energy in cells with a high local density (above 4.5 MIP equivalent per cell) and the total reconstructed energy for a subset of 20 GeV pion showers fully contained in the analog HCAL. a) Without the application of weighting, b) with weighting applied. Note that the selection cut of 4.5 MIP is applied on the unweighted energy density in both cases.}
\end{figure}

The weights were determined from a statistically independent data set with a chi-square minimization procedure, for 10 energy density bins, as shown for the HCAL in Figure \ref{fig:Binning}. The weights were then parametrized with a function with three free parameters, determined by an iterative fit procedure, to describe their energy dependence. Separate weight factors for each of the subdetectors (ECAL, HCAL, TCMT) were used. The reconstructed energy was also studied using the default case of one constant conversion factor per detector, without density dependent weighting. This reconstructed energy was also used to select the energy dependent weights, so no knowledge of the beam energy was necessary for the applications of the weighting procedure.

Figure \ref{fig:Correlation} a) shows the correlation of the total reconstructed energy in the HCAL for fully contained 20 GeV pion showers and the total energy in cells with a high local energy density (above 4.5 MIP equivalent per cell). It is clearly apparent that events with a high fraction of the total energy in cells with a high density tend also to have a high total energy, while those with a very low energy content above the threshold also have a lower reconstructed energy. The slope of the correlation is larger than unity, which is the basis of the weighting algorithm.

Figure \ref{fig:Correlation} b) shows the same correlation after the application of the density dependent weights. Since the procedure assigns a lower weight to cells with a high energy content, the impact of the energy in high density regions is significantly reduced. It is also apparent that the energy resolution, given by the width of the distribution of the total reconstructed energy, is significantly improved by the weighting algorithm.  

\begin{figure}[hbt]
\begin{center}
\includegraphics[width=0.48\textwidth]{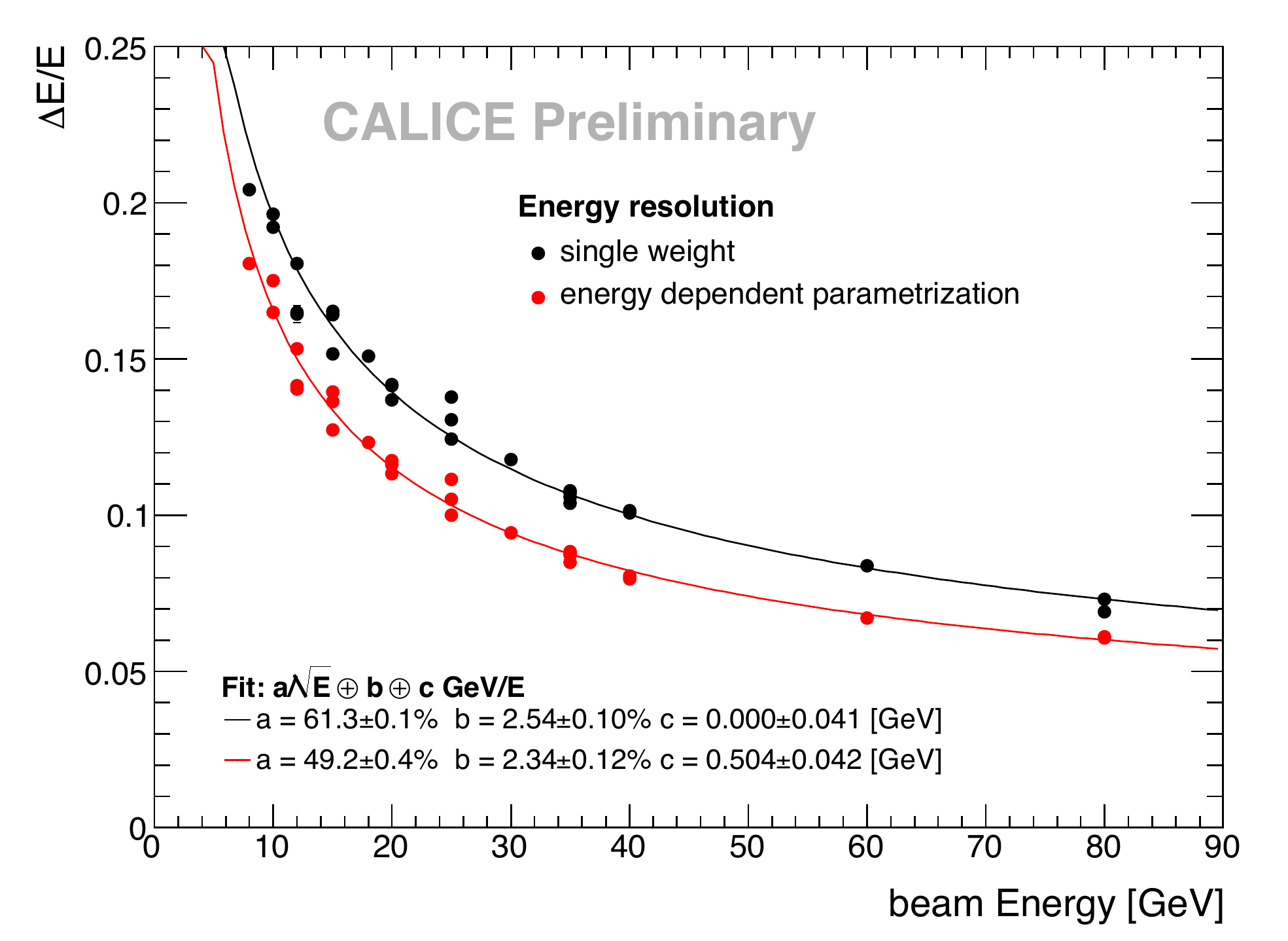}
\end{center}
\caption{\label{fig:Resolution} Energy resolution of the complete CALICE setup for hadrons at various energies as a function of beam energy, using energy reconstruction with one factor converting the corrected detector signal to energy, and with local energy density dependent weighting.}
\end{figure}

Figure \ref{fig:Resolution} shows the energy resolution for the complete \mbox{CALICE} setup, both for the reconstruction with one conversion factor per detector and for the density dependent weights. No requirements for shower containment or shower start positions were made. The reconstructed energy, and the corresponding energy resolution, for each data point was determined by a Gaussian fit to the energy distribution, in the range of $\pm1.5\, \sigma$ around the maximum value. It is apparent that the simple weighting method improves the energy resolution by about $20 \%$, yielding a stochastic term of $49.2\%/\sqrt{E\, [\mathrm{GeV}]}$.

\begin{figure}[hbt]
\begin{center}
\includegraphics[width=0.46\textwidth]{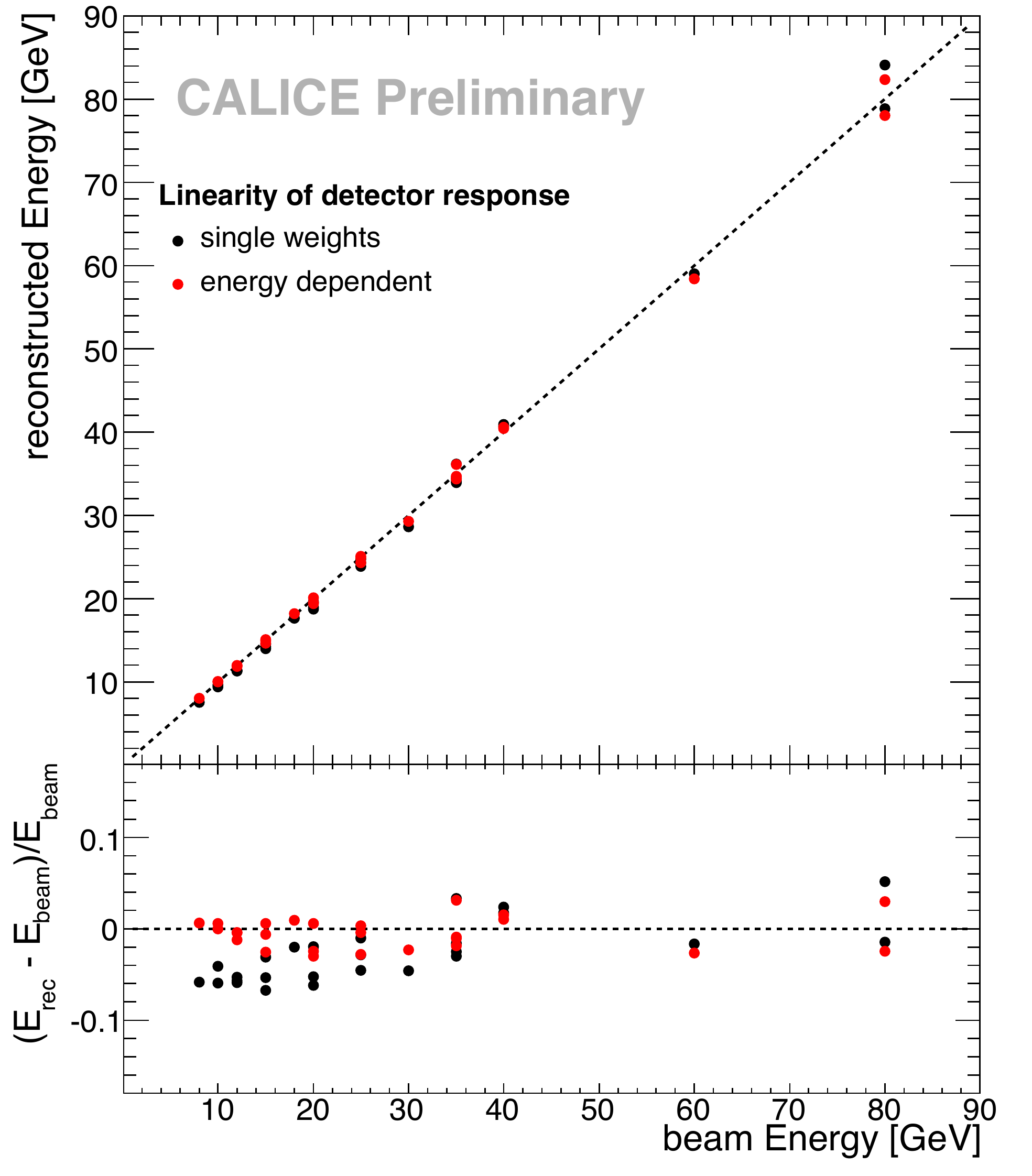}
\end{center}
\caption{\label{fig:Linearity}Reconstructed energy of the complete CALICE setup for hadrons at various energies as a function of beam energy, for reconstruction with one factor and with density dependent weighting. The lower panel shows the relative deviation of the reconstructed energy from the true beam energy.}
\end{figure}

Also the linearity of the detector response was improved, as shown in Figure \ref{fig:Linearity}. Over the full energy range from 8 GeV to 80 GeV, no significant non-linear behavior of the reconstructed energy is observed, with deviations of individual measurements of less than 4\% from the expected linear behavior. The spread of reconstructed energy and resolution for multiple measurements at the same beam energy is at least partially due to missing corrections for the temperature dependence of the detector response, which is currently being implemented into the analysis. 

Further improvement of the software compensation procedures might be possible by introducing a sub-clustering within the hadronic showers to directly identify electromagnetic subshowers. Combining larger volumes within the weighting procedure has the potential to reduce the influence of cell-to-cell fluctuations due to the extreme granularity of the detector.

\section{Summary}

The CALICE collaboration has constructed and tested highly granular `imaging' calorimeters for experiments at a future linear $e^+e^-$ collider. Detailed studies of calibration strategies and the required calibration precision of such detectors have led to a viable calibration procedure for the hadronic calorimeter in a complete detector system, using of-site calibration runs and in-situ module-by-module intercalibration using track segments identified within hadronic showers. 

\newpage
The high granularity of the hadron calorimeter, required for optimal performance with particle flow algorithms, also opens up the possibility for software compensation. A simple algorithm based on the weighting of the energy deposit in each individual cell according to the local energy density has been developed. This reconstruction method showed significantly improved linearity of the detector response, and an improvement of 20\% in the hadronic energy resolution, with a stochastic term of approximately 50\%/$\sqrt{E\, [\mathrm{GeV}]}$. Further improvements with more sophisticated algorithms seem possible.

\bibliographystyle{IEEEtran.bst}
\bibliography{CALICE}

\end{document}